\documentclass[twoside,twocolumn,english,pra,twocolumn,aps,superscriptaddress,showpacs]{revtex4-1}

\usepackage[T1]{fontenc}
\usepackage[latin9]{inputenc}
\usepackage{xcolor}
\usepackage{babel}
\usepackage{amsmath}
\usepackage{amsthm}
\usepackage{amssymb}
\usepackage{stmaryrd}
\usepackage{graphicx}
\usepackage[unicode=true,pdfusetitle,
 bookmarks=true,bookmarksnumbered=false,bookmarksopen=false,
 breaklinks=false,pdfborder={0 0 0},pdfborderstyle={},backref=false,colorlinks=true]
 {hyperref}

\makeatletter
\theoremstyle{plain}

\usepackage{tabularx}

\usepackage{makecell}
\usepackage{fontenc}

\hypersetup{urlcolor=blue}

\usepackage{cmap}
\usepackage[T1]{fontenc}

\input glyphtounicode
\makeatother

\begin{document}
\title{Phase estimation in lossy optical interferometry without a reference
beam}
\author{Jun Tang}
\altaffiliation{These authors contributed equally to this work}
\affiliation{Institute of Quantum Information and Technology, Nanjing University
of Posts and Telecommunications, Nanjing 210003, China}
\author{Dong-Qing Wang}
\altaffiliation{These authors contributed equally to this work}
\affiliation{Institute of Quantum Information and Technology, Nanjing University
of Posts and Telecommunications, Nanjing 210003, China}
\author{Wei Zhong}
\email{zhongwei1118@gmail.com}

\affiliation{Institute of Quantum Information and Technology, Nanjing University
of Posts and Telecommunications, Nanjing 210003, China}
\author{Lan Zhou}
\affiliation{School of Science, Nanjing University of Posts and Telecommunications,
Nanjing 210003, China}
\author{Yu-Bo Sheng}
\affiliation{College of Electronic and Optical Engineering, Nanjing University
of Posts and Telecommunications, Nanjing 210003, China China}
\begin{abstract}
We investigate phase estimation in a lossy interferometer using entangled
coherent states, with particular focus on a scenario where
no reference beam is employed. By calculating the quantum Fisher information, 
we reveal two key results: (1) the metrological equivalence
between scenarios with and without a reference beam, established under
ideal lossless conditions for the two-phase-shifting configuration,
breaks down in the presence of photon loss, and (2) the pronounced
inferior performance of entangled coherent states relative to NOON states, observed in
the presence of a reference beam, disappears in its absence.
\end{abstract}
\maketitle

\section{Introduction}

Optical interferometers are among the most precise measurement instruments
and have been widely employed in diverse fields, such as gravitational
wave detection \citep{Tse2019PRL,Acernese2019PRL}, quantum imaging
\citep{Defienne2024review,Jin2024review}, quantum ranging \citep{Qian2023PRL}
and quantum lithography \citep{DAngelo2001PRL}. The ultimate sensitivity
for estimating an unknown phase in an interferometer is typically
determined by the state of input light. When classical light is used,
the sensitivity is bounded by the shot noise limit $1/\sqrt{N}$,
where $N$ is the average photon number \citep{Giovannetti2004,Giovannetti2011,Giovannetti2006PRL}.
In contrast, non-classical states of light can surpass this limit.
Among them, NOON states are particularly renowned for their ability
to achieve Heisenberg-limited sensitivity, which scales as $1/N$
\citep{Giovannetti2004,Giovannetti2011,Giovannetti2006PRL}. More
recently, entangled coherent states (ECSs) have emerged as promising candidates for phase estimation
as they not only surpass the Heisenberg limit for small $N$ but also
exhibit greater robustness to photon loss compared to NOON states
\citep{Luis2001PRA,Joo2011PRL,Joo2012PRA,Zhang2013PRA,Jing2014bCTP,Zhong2023QINP,Tang2025CPB,Yu2018OE}.

To fully leverage the advantages of non-classical light, an effective
and sensitive measurement is required \citep{Braunstein1994PRL,Braunstein1996AP,PARIS2009review,Zhong2014JPA}.
In practical implementations, photon-number-resolving detectors constitute
a crucial class of measurement schemes, such as parity and photon
count \citep{Campos2003,Hofmann2009PRA,Plick2010,Chiruvelli2011,Seshadreesan2011,Seshadreesan2013PRA,Pezze2013PRL,Zhong2017PRA,Thekkadath2020npj,Zhong2021PRA}.
These detectors are experimentally favorable, as they can be implemented
without the need for a shared reference beam. However, previous studies
on phase estimation with ECSs have been conducted under assumption
that a common reference beam is already established \citep{Luis2001PRA,Joo2011PRL,Joo2012PRA,Zhang2013PRA,Jing2014bCTP,Yu2018OE}.
As a result, the findings reported in these works are generally not
applicable to measurement schemes involving photon-number-resolving
detectors. This naturally raises the question: how do ECSs perform
in phase estimation when the reference beam is absent? More specifically,
what is the ultimate phase sensitivity achievable for ECSs when using
photon-number resolving detectors?

In this manuscript, we address this issue by reexamining the metrological
performance of ECSs in a lossy interferometer. Using the quantum Fisher information
(QFI) framework \citep{Braunstein1994PRL,Braunstein1996AP,PARIS2009review,Zhong2014JPA},
we evaluate the ultimate sensitivity for both scenarios with and without
a reference beam. Although it is commonly acknowledged that these
two scenarios yield equivalent sensitivities under ideal lossless
conditions for the two-phase-shifting configuration \citep{Jarzyna2012PRA,Zhong2023QINP},
our results show that significant differences arise in the presence
of photon loss. Specifically, not only does the equivalence break
down, but also the disadvantage of ECSs relative to NOON states observed
in the presence of reference beam also disappears when the reference
beam is omitted.

This paper is organized as follows. In Sec. II, we introduce the two-mode
optical interferometer and review the fundamentals of quantum estimation
theory. Sec. III provides a comprehensive comparison of phase sensitivities
for ECSs with and without a reference beam. Finally, we conclude in
Sec. IV.

\section{phase estimation with a two-mode optical interferometer}

\begin{figure}[t]
\centering{}\includegraphics[scale=0.25]{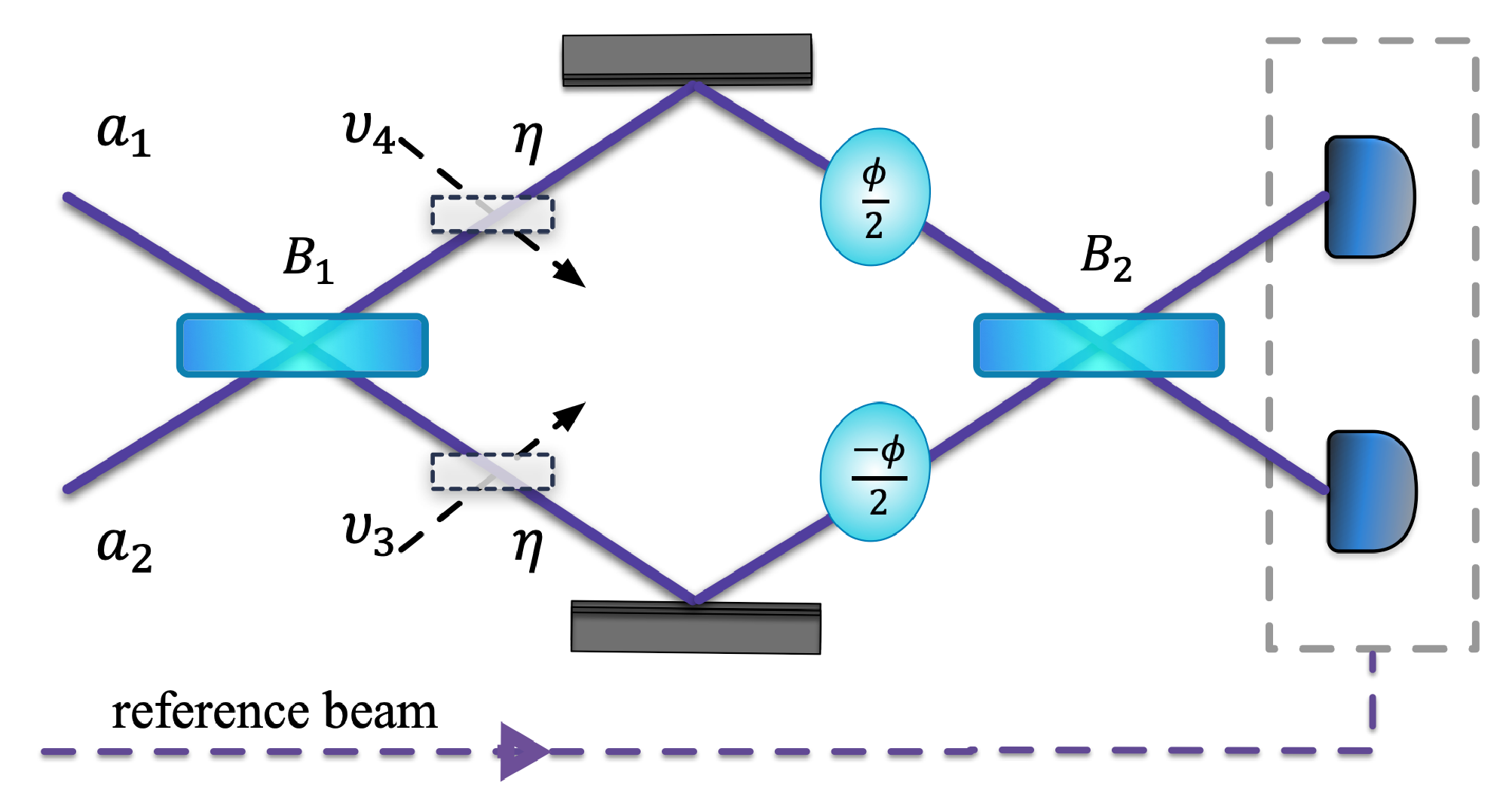}\caption{Schematic of a two-mode lossy optical interferometer. \label{fig:MZI}}
\end{figure}

A two-mode optical interferometer enables precise measurement of the
phase difference between the two paths (see Fig.~\ref{fig:MZI}).
A typical interferometer comprises two balanced beam splitters (BSs)
$B_{i}\left(i=1,2\right)$ and a phase shifter $U_{\phi}$ with an
unknown phase parameter $\phi$ \citep{Yurke1986PRA}. As photons
propagate between the BSs, the phase of interest is accumulated. The
overall interferometric evolution can be described by the composite
operator $K\!=\!B_{2}U_{\phi}B_{1}$. If $\rho_{{\rm in}}$ denotes
the state entering the interferometer, then the output state is given
by $\rho_{{\rm out}}\!=\!K\rho_{{\rm in}}K^{\dagger}$. Measurements
performed at the output ports provide information to estimate the
unknown phase parameter.

For convenience, we refer to the probe state as the state prior to
the phase-shifting operation, i.e., $\rho\!=\!B_{1}\rho_{{\rm in}}B_{1}^{\dagger}$.
Under the action of the phase shifter $U_{\phi}$, the state evolves
into the phase-encoded state $\rho_{\phi}\!=\!U_{\phi}\rho U_{\phi}^{\dagger}$.
According to quantum estimation theory, the quantum Cram\'er-Rao
theorem sets a fundamental lower bound on the phase uncertainty $\delta\hat{\phi}$
for any locally unbiased estimator $\hat{\phi}$ \citep{Helstrom1976Book,Holevo1982Book,Braunstein1994PRL}:

\begin{eqnarray}
\delta\hat{\phi} & \geqslant & \left(mF\right)^{-1/2},
\end{eqnarray}
where $m$ is the number of repetitions of an experiment, and $F$
is the QFI defined as $F\!=\!{\rm Tr}\left(\rho_{\phi}L^{2}\right)$.
Here, $L$ is the symmetric logarithmic derivative operator implicitly
defined by $\partial\rho_{\phi}/\partial\phi\!=\!\left(\rho_{\phi}L+L\rho_{\phi}\right)/2$.
This bound is asymptotically achievable and serves as a benchmark
for assessing the performance of phase estimation protocols. In interferometric
phase estimation, the QFI depends solely on the phase-encoded state
$\rho_{\phi}$, regardless of the second BS $B_{2}$ due to the $\phi$-independent
unitary invariance of the QFI \citep{Liu2019JPA}.

We consider two commonly used forms of phase-shifting unitary operators.
The first is the two-arm phase shift
\begin{equation}
U_{\phi}^{T}=\text{exp}\left[-i\phi\left(a_{1}^{\dagger}a_{1}-a_{2}^{\dagger}a_{2}\right)/2\right],\label{eq:phaseshifting}
\end{equation}
which in introduce a difference phase shift by applying phase shifts
of $\phi/2$ and $-\phi/2$ to the two interferometer paths, respectively.
Here, $a_{i}$ and $a_{i}^{\dagger}$ denote the annihilation and
creation operators for the $i$th mode ($i=1,2$), respectively. The
second is the single-arm phase shift $U_{\phi}^{S}\!=\!e^{-i\phi a_{1}^{\dagger}a_{1}}$,
which applies the full phase shift $\phi$ to a single path. Notably,
these two phase-shifting operators are metrologically equivalent in
the absence of reference beam \citep{Jarzyna2012PRA,Zhong2023QINP}.
This equivalence stems from the fact that $U_{\phi}^{S}$ differs
from $U_{\phi}^{T}$ only up to a sum phase shift $U_{\phi}^{\binampersand}\!=\!\text{exp}[-i\phi(a_{1}^{\dagger}a_{1}\!+\!a_{2}^{\dagger}a_{2})/2]$,
which is experimentally immeasurable without introducing an external
phase reference.

More precisely, in reference-free scenarios, the probe state must
be phase-averaged as \citep{Jarzyna2012PRA,Bartlett2007RMP}
\begin{eqnarray}
\varrho & = & \int_{-\pi}^{\pi}\frac{d\theta}{2\pi}U_{\theta}^{a_{1}}U_{\theta}^{a_{2}}\rho U_{\theta}^{a_{1}\dagger}U_{\theta}^{a_{2}\dagger},\label{eq:phaseaveraging}
\end{eqnarray}
with $U_{\theta}^{x}\!=\!\exp(-i\theta x^{\dagger}x)$. The resulting
phase-averaged state is a statistical ensemble of states with fixed
photon numbers, resulting in the loss of coherence between different
photon-number subspaces. Consequently, such states are insensitive
to $U_{\phi}^{\binampersand}$, rendering $U_{\phi}^{S}$ and $U_{\phi}^{T}$
operationally indistinguishable in phase estimation. In other words,
for given a probe state, the ultimate phase sensitivity is independent
of the specific form of the phase-shifting operation in the absence
of a reference beam. However, once a external reference beam is established,
the sum phase becomes physically meaningful, and the two configurations
$U_{\phi}^{S}$ and $U_{\phi}^{T}$ become distinguishable, leading
to potentially different metrological performances.

In this work, we focus on the $U_{\phi}^{T}$ configuration for the
following reasons: (1) In the ideal lossless case, the QFI under $U_{\phi}^{T}$
is identical irrespective of the presence or absence of a reference
beam. However, whether this equivalence persists under photon loss
remains an open question, which is one of issues we aim to address.
(2) The $U_{\phi}^{S}$ configuration has been extensively studied
in previous studies \citep{Joo2011PRL,Joo2012PRA,Jing2014bCTP,Zhang2013PRA,Tang2025CPB},
and the methodology developed here for $U_{\phi}^{T}$ can be straightforwardly
adapted to analyze $U_{\phi}^{S}$ as well.

\section{Phase sensitivity of lossy interferometry with ECSs}

In this section, we investigate the phase sensitivity of a lossy interferometer
using ECSs as the probe state. Photon loss is modeled by inserting
a virtual BS with transmittance $\eta$ into each interferometer path,
denoted by the operator $V_{\eta}$ \citep{Dorner2009PRL,Demkowicz-Dobrzanski2009PRA,Joo2012PRA}.
For simplicity, we assume equal photon loss in both interferometer
paths. ECSs can be generated by mixing coherent and coherent superposition
states of light on a BS \citep{Luis2001PRA}, or alternatively generated
by mixing coherent and squeezed vacuum states of light on a BS \citep{Israel2019Optica}.
The resulting ECS is given by
\begin{eqnarray}
\left|{\rm ECS}\right\rangle  & = & \mathcal{N}\!\left(\vert\alpha\rangle\vert0\rangle+\vert0\rangle\vert\alpha\rangle\right),\label{eq:ECS}
\end{eqnarray}
with normalization coefficient $\mathcal{N}\!=\!1/\!\sqrt{2(1+e^{-\left|\alpha\right|^{2}})}$.
This state can be expanded as a superposition of NOON states
\begin{eqnarray}
\left|{\rm ECS}\right\rangle  & = & \sqrt{2}\mathcal{N}\sum_{n=0}^{\infty}|c_{n}|^{2}\vert n\!::\!0\rangle,
\end{eqnarray}
where $c_{n}\!=\!e^{-\vert\alpha\vert^{2}\!/2}\alpha^{n}/\sqrt{n!}$,
and $\vert n\negthickspace::\negthickspace0\rangle\!\equiv\!(\vert n\rangle\vert0\rangle+\vert0\rangle\vert n\rangle)/\sqrt{2}$
denotes a NOON state with fixed photon number $n$. The average photon
number of the ECS is $\overline{N}\!=\!2\mathcal{N}^{2}|\alpha|^{2}$.
In the limit of large $\left|\alpha\right|$, this approaches $\overline{N}\!\sim\!|\alpha|^{2}$
since $\mathcal{N}\!\sim\!1/\sqrt{2}$. In what follows, we compute
the QFI for ECS-based phase estimation within two distinct scenarios:
with and without a reference beam.

\subsection{Phase sensitivity without a reference beam}

We first consider the scenario in which no reference beam is available.
In this case, the phase-averaging operation defined in Eq.~\eqref{eq:phaseaveraging}
must be applied. As a result, the ECS probe state given in Eq.~\eqref{eq:ECS}
becomes a mixed state, which can be expressed as a direct sum of weighted
NOON states \citep{Zhong2023QINP,Liu2019JPA}
\begin{eqnarray}
\varrho_{{\rm ECS}} & = & 2\mathcal{N}^{2}\bigoplus_{n=0}^{\infty}\left|c_{n}\right|^{2}\left|n\!::\!0\right\rangle \!\left\langle n\!::\!0\right|.\label{eq:mxiedECS}
\end{eqnarray}
According to the additivity property of the QFI, the QFI for this
phase-averaged ECS can be directly calculated as

\begin{eqnarray}
F_{\!\varrho} & = & 2\mathcal{N}^{2}\!\sum_{n=0}^{\infty}\left|c_{n}\right|^{2}F_{{\rm noon}},\label{eq:QFI-mECS}
\end{eqnarray}
where $F_{{\rm noon}}\!=\!n^{2}\eta^{n}$ is the QFI for small NOON
states used as probe states in a lossy interferometer \citep{Zhong2021PRA,Zhang2013PRA}.
Equation~\eqref{eq:QFI-mECS} can be expressed in the compact form
\begin{eqnarray}
F_{\!\varrho} & = & 2\mathcal{N}^{2}e^{-\left|\alpha\right|^{2}\left(1-\eta\right)}\!\left(\left|\alpha\right|^{4}\eta^{2}+\left|\alpha\right|^{2}\eta\right).\label{eq:QFI_mECS_simple}
\end{eqnarray}
In this expression, the first term inside the parentheses represents
the Heisenberg-scaling contribution, while the second term corresponds
to shot-noise scaling. This result is valid for both the single-phase
and two-phase configurations ($U_{\phi}^{S}$ and $U_{\phi}^{T}$),
as justified in preceding section. In the ideal lossless case $(\eta\!=\!1)$
, the QFI simplifies to \citep{Zhong2023QINP}
\begin{eqnarray}
F_{\!\varrho} & = & 2\mathcal{N}^{2}\!\left(|\alpha|^{4}+|\alpha|^{2}\right).\label{eq:QFI_pure}
\end{eqnarray}
Expressing in terms of mean photon number, we have $F_{\!\varrho}\geq\overline{N}^{2}+\overline{N}$,
thereby surpassing the conventional Heisenberg limit. This result
demonstrates that ECSs offer superior phase sensitivity compared to
NOON states with the same average photon number.

\subsection{Phase sensitivity with a reference beam}

For comparison, we now consider the scenario in which a common reference
beam is available, and the probe state is the pure ECS defined in
Eq.~\eqref{eq:ECS}. In this case, the phase-shifting operation is
implemented using the operator $U_{\phi}^{T}$. Unlike the reference-free
scenario, calculating the QFI in the presence of a reference beam
is more intricate. Below, we summarize the key steps
in the calculation, while full derivations are provided in Appendix.

Owing to the commutation relationship between photon loss and phase
shifting \citep{Dorner2009PRL,Demkowicz-Dobrzanski2009PRA}, the order
of these operations can be interchanged without affecting the final
measurement results. Thus, we assume that the ECS in Eq.~\eqref{eq:ECS}
first undergoes photon loss, followed by the phase accumulation process.
Under such loss, the ECS evolves into a mixed state as
\begin{eqnarray}
\sigma_{{\rm ECS}} & = & {\rm Tr}_{34}\!\left[V_{13}V_{24}\left(\vert{\rm ECS}\rangle_{12}\langle{\rm ECS}\vert\!\otimes\!\vert00\rangle_{34}\langle00\vert\right)V_{24}^{\dagger}V_{13}^{\dagger}\right]\nonumber \\
 & = & \mathcal{N}^{2}\Big(\vert\alpha\sqrt{\eta},0\rangle\langle\alpha\sqrt{\eta},0\vert+\vert0,\alpha\sqrt{\eta}\rangle\langle0,\alpha\sqrt{\eta}\vert\nonumber \\
 &  & +e^{-\left(1-\eta\right)\left|\alpha\right|^{2}}\vert\alpha\sqrt{\eta},0\rangle\langle0,\alpha\sqrt{\eta}\vert\nonumber \\
 &  & +e^{-\left(1-\eta\right)\left|\alpha\right|^{2}}\vert0,\alpha\sqrt{\eta}\rangle\langle\alpha\sqrt{\eta},0\vert\Big),\label{eq:lossy_ECS}
\end{eqnarray}
where $\vert0\rangle_{k}\left(k\!=\!3,4\right)$ denote the vacuum
states of the environmental modes corresponding to paths $1$ and
$2$, respectively. Here the virtual beam splitters are defined as
$V_{13}$ and $V_{24}$ define $V_{13}\!=\!\exp[\arccos\!\sqrt{\eta}(a_{1}^{\dagger}\upsilon_{3}-a_{1}\upsilon_{3}^{\dagger})]$,
and $V_{24}$ is defined analogously by substituting modes $1$ and
$3$ with modes $2$ and $4$. Let $\vert\varPsi_{1}\rangle\!=\!\vert\alpha\sqrt{\eta},0\rangle$
and $\vert\varPsi_{2}\rangle\!=\!\vert0,\alpha\sqrt{\eta}\rangle$.
These states are non-orthogonal, with overlap $p\!\equiv\!\langle\varPsi_{1}\vert\varPsi_{2}\rangle\!=\!e^{-\eta\left|\alpha\right|^{2}}$.
Employing the Gram-Schmidt orthogonalization and performing spectral
decomposition \citep{Jing2014bCTP,Tang2025CPB}, the state $\sigma_{{\rm ECS}}$
can be diagonalized as
\begin{eqnarray}
\sigma_{{\rm ECS}} & = & \gamma_{+}\vert\gamma_{+}\rangle\langle\gamma_{+}\vert+\gamma_{-}\vert\gamma_{-}\rangle\langle\gamma_{-}\vert,\label{eq:lossy_ECS_diagonal}
\end{eqnarray}
where the eigenstates take the form
\begin{equation}
\left|\gamma_{\pm}\right\rangle =\mathcal{C}_{\pm}\left|\varPsi_{1}\right\rangle +\mathcal{D}_{\mp}\left|\varPsi_{2}\right\rangle ,\label{eq:eigenvectors}
\end{equation}
and the corresponding eigenvalues are given by
\begin{equation}
\gamma_{\pm}=\frac{1}{2}\left(1\pm\sqrt{1-\det\!\sigma_{{\rm ECS}}}\right).
\end{equation}
In Eq.~\eqref{eq:eigenvectors}, the expansion coefficients are defined
as
\begin{equation}
\mathcal{C}_{\pm}=\pm\zeta_{\pm}-p\mathcal{D}_{\mp},\;\mathcal{D}_{\mp}=\frac{\zeta_{\mp}}{\sqrt{1-p^{2}}},
\end{equation}
with $\zeta_{\pm}\!=\!\sqrt{\frac{\sqrt{1-4\det\!\sigma_{{\rm ECS}}}\pm\langle\sigma_{3}\rangle}{2\sqrt{1-4\det\!\sigma_{{\rm ECS}}}}},$
$\det\!\sigma_{{\rm ECS}}\!=\!\mathcal{N}^{4}(1\!-\!p^{2})(1\!-\!p_{\bot}^{2})$,
$\langle\sigma_{3}\rangle\!=\!1\!-\!2\mathcal{N}^{2}(1\!-\!p^{2})$
and $p_{\bot}\!=\!e^{-(1-\eta)\left|\alpha\right|^{2}}$.

The QFI for the mixed state $\sigma_{{\rm ECS}}$ is then obtained
by

\begin{eqnarray}
F_{\!\sigma} & = & 4\!\left(\gamma_{+}\Delta^{2}G_{+}+\gamma_{-}\Delta^{2}G_{-}-4\gamma_{+}\gamma_{-}\!\left|G_{+-}\right|^{2}\right),\quad\label{eq:QFI_nonorthogonal}
\end{eqnarray}
where
\begin{eqnarray}
\Delta^{2}G_{\pm} & = & \left\langle \gamma_{\pm}\right|G^{2}\left|\gamma_{\pm}\right\rangle -\left\langle \gamma_{\pm}\right|G\left|\gamma_{\pm}\right\rangle ^{2},\\
G_{+-} & = & \left\langle \gamma_{+}\right|G\left|\gamma_{-}\right\rangle ,
\end{eqnarray}
and $G\!=\!(a_{1}^{\dagger}a_{1}\!-\!a_{2}^{\dagger}a_{2})/2$ is
the generator of $U_{\phi}^{T}$ defined in Eq.~\eqref{eq:phaseshifting}.
To compute $\Delta^{2}G_{\pm}$ and $G_{+-}$, we use the following
expectation values

\begin{eqnarray}
\left\langle \varPsi_{1}\right|G\left|\varPsi_{1}\right\rangle  & \!=\! & \frac{1}{2}\left|\alpha\right|^{2}\!\eta,\quad\left\langle \varPsi_{2}\right|G\left|\varPsi_{2}\right\rangle \!=\!-\frac{1}{2}\left|\alpha\right|^{2}\!\eta,\\
\left\langle \varPsi_{i}\right|G^{2}\left|\varPsi_{i}\right\rangle  & \!=\! & \frac{1}{4}\left(\left|\alpha\right|^{2}\!\eta+\left|\alpha\right|^{4}\!\eta^{2}\right)\!,{\rm for\;}i=1,2,\\
\left\langle \varPsi_{1}\right|G\left|\varPsi_{2}\right\rangle  & \!=\! & \left\langle \varPsi_{1}\right|G^{2}\left|\varPsi_{2}\right\rangle \!=\!0.
\end{eqnarray}
Substituting these into Eq.~\eqref{eq:QFI_nonorthogonal} yields
the full expression for $F_{\!\sigma}$. Although this expression
is algebraically cumbersome, it simplifies, in the limit $p\!\rightarrow\!0$
(i.e., $\eta|\alpha|^{2}\!\gg\!1$), to

\begin{eqnarray}
F_{\!\sigma} & = & 2\mathcal{N}^{2}\!\left(e^{-2\left|\alpha\right|^{2}\left(1-\eta\right)}\!\left|\alpha\right|^{4}\eta^{2}+\left|\alpha\right|^{2}\eta\right).\label{eq:QFI_lossyECS_simple}
\end{eqnarray}
In the ideal lossless case ($\eta\!=\!1$), this expression reduces
to Eq.~\eqref{eq:QFI_pure}, i.e., $F_{\!\varrho}\!=\!F_{\!\sigma}$,
thereby confirming that, for the two-phase-shifting operation $U_{\phi}^{T}$,
the metrological performance remains invariant regardless of the presence
or absence of a reference beam. This result is consistent with the
commonly acknowledged conclusion reported in \citep{Jarzyna2012PRA,Zhong2023QINP},
namely that the phase shifter $U_{\phi}^{T}$ exhibits metrological
equivalence between scenarios with and without a reference beam for
phase estimation protocols employing pure probe states. However, as
we demonstrate below, in the presence of photon loss, ECSs with and
without a reference beam exhibit distinct metrological behavior, indicating
that this equivalence does not extend to lossy conditions.

\subsection{Further comparison}

To thoroughly assess metrological performance under photon loss, we
compare three quantum strategies: (i) ECSs without a reference beam,
(ii) ECSs with a reference beam, and (iii) NOON states, whose QFI
is given by $F_{{\rm NOON}}\!=\!\eta^{N}N^{2}$ \citep{Zhong2021PRA,Zhang2013PRA}.
For a fair comparison, we set the same mean photon
number $\overline{N}=N$ for both ECSs and NOON states.
\begin{figure}[t]
\begin{centering}
\includegraphics[scale=0.245]{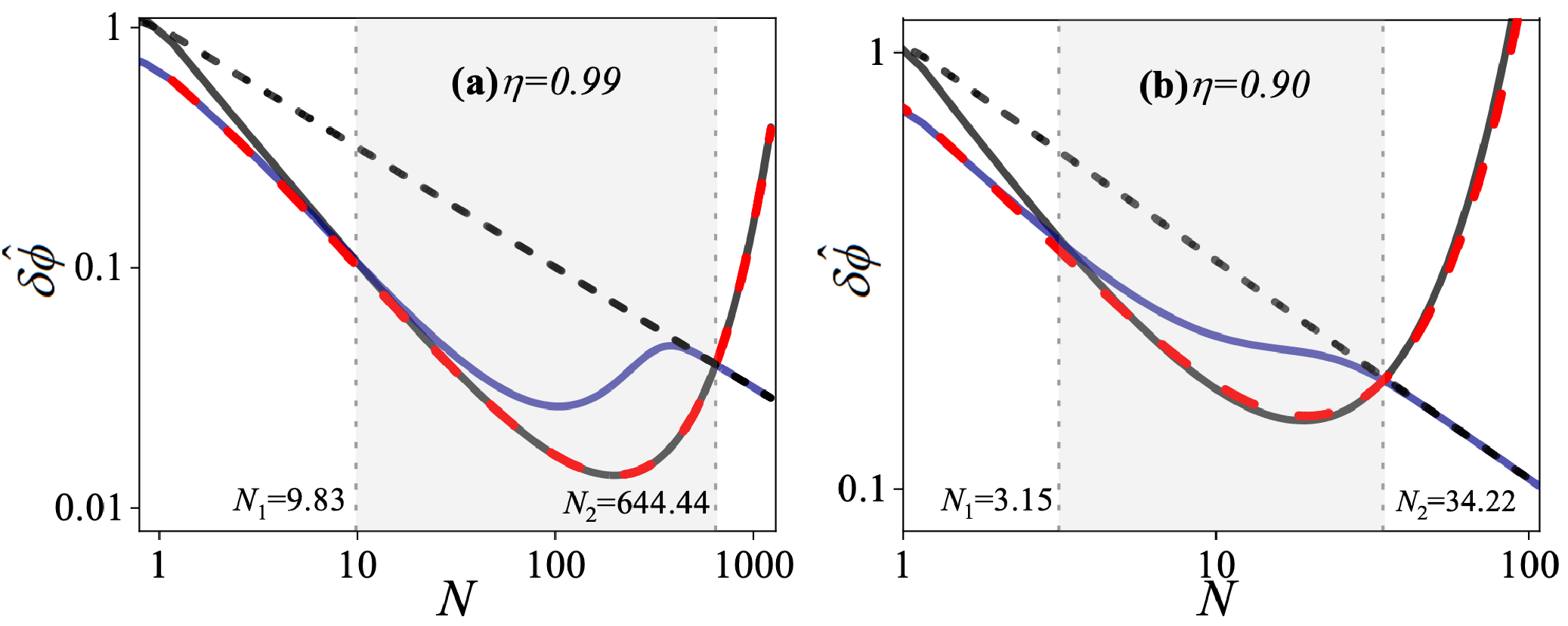}
\par\end{centering}
\caption{Log-log plot of phase sensitivity as a function of the mean photon
number $N$ (with $\overline{N}=N$ for ECSs) for
(a) $\eta\!=\!0.99$ and (b) $\eta\!=\!0.9$. The blue solid line
and red dashed line represent ECSs with and without a reference beam,
respectively. The black solid line corresponds to the NOON states,
and the gray dashed line marks the shot-noise limit $1/\sqrt{\eta N}$
as a benchmark. \label{fig:phase_sensitivity}}
\end{figure}

Let us first compare the strategies of NOON states and ECSs with a
reference beam. As shown in Fig.~\eqref{fig:phase_sensitivity},
the two critical crossing points $N_{1}$ and $N_{2}$ exist, defined
by the condition $F_{{\rm NOON}}\!=\!F_{\!\sigma}$. Although their
analytical expressions are cumbersome, these crossing points demarcate
distinct performance regimes. In the regime $N\!<\!N_{1}$, ECSs with
a reference beam outperforms NOON states. However, in the intermediate
range $N_{1}\!<\!N\!<\!N_{2}$, this trend reverses with NOON states
exhibiting superior performance against ECSs. As $N$ increases further,
the QFI of NOON states diverges, whereas the QFI of ECSs with a reference
beam asymptotically approaches the shot-noise limit ($\delta\phi\!\sim\!1/\sqrt{\eta N}$).
This phenomenon illustrates that although the use of a reference beam
enables ECSs to achieve superior sensitivity compared to NOON states
at low $N$, this advantage diminishes at intermediate $N$. A similar
trend has been reported for the single-phase configuration $U_{\phi}^{S}$
\citep{Zhang2013PRA}. Interestingly, ECSs without a reference beam
combines the advantages of both strategies: for $N\!<\!N_{1}$, they
perform similarly to ECSs with a reference beam, while for $N\!>\!N_{1}$,
they resemble NOON states. Consequently, although phase sensitivity
deteriorates with increasing loss, omitting the reference beam can
be advantageous in ECS-based phase estimation, offering improved performance
across a border range of photon numbers.

These behaviors are quantitatively supported by Eqs.~\eqref{eq:QFI_mECS_simple}
and \eqref{eq:QFI_lossyECS_simple}. In the low intensity regime,
where $\left|\alpha\right|^{2}\!\left(1-\eta\right)\!\ll\!1$, we
have $e^{-2\left|\alpha\right|^{2}\left(1-\eta\right)}\!\sim\!1$,
which leads to
\begin{eqnarray}
F_{\!\varrho} & = & F_{\!\sigma}\approx2\mathcal{N}^{2}\left(\left|\alpha\right|^{4}\!\eta^{2}+\left|\alpha\right|^{2}\!\eta\right).
\end{eqnarray}
This explains why, in Fig.~\eqref{fig:phase_sensitivity}, the QFIs
of both ECS strategies nearly coincide in the low-photon-number regime.
In the high intensity regime $\overline{N}\!\sim\!\left|\alpha\right|^{2}\!\gg\!1$,
however, the behavior diverges. For ECSs with a reference beam, the
first term in Eq.~\eqref{eq:QFI_lossyECS_simple} (which corresponds
to the Heisenberg-limit scaling) vanishes due to $e^{-2\left|\alpha\right|^{2}\left(1-\eta\right)}\!\rightarrow\!0$.
This leaves only the second term, which scales as shot noise limit:
$F_{\!\sigma}\!\sim\!\eta\overline{N}$. By contrast, for ECSs without
a reference beam, the behavior resembles that of NOON states in this
regime. Recalling that $\overline{N}\!=\!\left|\alpha\right|^{2}$,
the QFI from Eq.~\eqref{eq:QFI_mECS_simple} becomes
\begin{eqnarray}
F_{\!\varrho} & = & e^{\left(\overline{N}+2\right)\left(\eta-1\right)}\overline{N}^{2}+e^{\overline{N}\left(\eta-1\right)}\overline{N}\eta\approx e^{\overline{N}\left(\eta-1\right)}\overline{N}^{2}.
\end{eqnarray}
This scaling is consistent with that of NOON states, for which
\begin{equation}
F_{{\rm NOON}}=\eta^{N}N^{2}\approx e^{N\left(\eta-1\right)}N^{2},
\end{equation}
where we have used the approximation $\ln\eta\!\sim\!\eta\!-\!1$
in the limit $\eta\!\rightarrow\!1$.

\section{Conclusion \label{sec:Conclusion}}

In this work, we systematically analyzed the phase estimation performance
of a lossy optical interferometer using ECSs as probe states, considering
scenarios both with and without a reference beam. Using the QFI as
a metric, we demonstrated that photon loss breaks the metrological
equivalence between these two scenarios in the two-phase-shifting
configuration. Furthermore, we showed that omitting the reference
beam may be beneficial for improving the sensitivity of ECS-based
interferometric phase estimation. These findings provide valuable
insights into lossy interferometry and the development of practical
quantum metrology schemes.

\section*{Acknowledgments}

This work was supported by the NSFC through Grant No. 12005106 and
the Postgraduate Research and Practice Innovation Program of Jiangsu
Province (Grant No. JSCX23-0260).

\section*{Appendix: Detailed Derivations\label{sec:AppendixA}}

In this appendix, we present the detailed derivations
of Eqs.~\eqref{eq:lossy_ECS}, \eqref{eq:lossy_ECS_diagonal} and
\eqref{eq:QFI_nonorthogonal} from the main text.

\subsection{Derivation of Eq.~\eqref{eq:lossy_ECS}}

\makeatletter
\renewcommand{\theequation}{A\arabic{equation}}
\makeatother
\setcounter{equation}{0}
For a lossy interferometer, photon losses are modeled
by inserting two virtual BSs $V_{13}$ and $V_{24}$ into each interferometer
path. The BS $V_{13}$ is defined as
\begin{eqnarray}
V_{13} & = & \exp\left[\arccos\!\sqrt{\eta}\left(a_{1}^{\dagger}\upsilon_{3}-a_{1}\upsilon_{3}^{\dagger}\right)\right],\label{eq:BS13}
\end{eqnarray}
where the subscripts $1$ and $3$ represent the interferometer path
$1$ and its corresponding environment mode $3$, which is initially
in the vacuum state. Similarly, $V_{24}$ acts on modes $2$ and $4$
in the same form, with $1$ and $3$ replaced by $2$ and $4$. Both
BSs have equal transmittance $\eta$. The input and output relations
for $V_{13}$ are
\begin{eqnarray}
V_{13}^{\dagger}a_{1}V_{13} & = & \sqrt{\eta}a_{1}+\sqrt{1-\eta}\upsilon_{3},\\
V_{13}^{\dagger}\upsilon_{3}V_{13} & = & -\sqrt{1-\eta}a_{1}+\sqrt{\eta}\upsilon_{3}.
\end{eqnarray}
Applying these, one obtains
\begin{eqnarray}
V_{13}\vert\alpha\rangle_{1}\vert0\rangle_{3} & = & \vert\alpha\sqrt{\eta}\rangle_{1}\,\vert\!-\!\sqrt{1\!-\eta}\alpha\rangle_{3}.
\end{eqnarray}
and similarly
\begin{eqnarray}
V_{24}\vert\alpha\rangle_{2}\vert0\rangle_{4} & = & \vert\alpha\sqrt{\eta}\rangle_{2}\,\vert\!-\!\sqrt{1\!-\eta}\alpha\rangle_{4}.
\end{eqnarray}
The total state after loss becomes
\begin{align}
V_{1,3}V_{2,4} & \left|{\rm ECS}\right\rangle _{12}\!\vert00\rangle_{3,4}\nonumber \\
 & =V_{1,3}V_{2,4}\mathcal{N}\!\left(\vert\alpha\rangle_{1}\vert0\rangle_{2}+\vert0\rangle_{1}\vert\alpha\rangle_{2}\right)\!\vert00\rangle_{3,4}\nonumber \\
 & =\mathcal{N}\Big(\!\vert\alpha\sqrt{\eta}\rangle_{1}\vert0\rangle_{2}\,\vert\!-\!\sqrt{1\!-\eta}\alpha\rangle_{3}\vert0\rangle_{4}\nonumber \\
 & \quad\,+\!\vert0\rangle_{1}\vert\alpha\sqrt{\eta}\rangle_{2}\vert0\rangle_{3}\,\vert-\sqrt{1\!-\eta}\alpha\rangle_{4}\!\Big).
\end{align}
Tracing over the environment modes $3$ and $4$ yields the reduced
density matrix for mode $1$ and $2$

\begin{eqnarray}
\sigma_{{\rm ECS}} & = & {\rm Tr}_{34}\!\left[V_{13}V_{24}\left(\vert{\rm ECS}\rangle_{12}\langle{\rm ECS}\vert\!\otimes\!\vert00\rangle_{34}\langle00\vert\right)V_{24}^{\dagger}V_{13}^{\dagger}\right]\nonumber \\
 & = & \mathcal{N}^{2}\Big(\vert\alpha\sqrt{\eta},0\rangle_{12}\langle\alpha\sqrt{\eta},0\vert+\vert0,\alpha\sqrt{\eta}\rangle_{12}\langle0,\alpha\sqrt{\eta}\vert\nonumber \\
 &  & +e^{-\left(1-\eta\right)\left|\alpha\right|^{2}}\vert\alpha\sqrt{\eta},0\rangle_{12}\langle0,\alpha\sqrt{\eta}\vert\nonumber \\
 &  & +e^{-\left(1-\eta\right)\left|\alpha\right|^{2}}\vert0,\alpha\sqrt{\eta}\rangle_{12}\langle\alpha\sqrt{\eta},0\vert\Big),\label{eq:cgm_ECS}
\end{eqnarray}
where we have used the overlap of coherent states
\begin{eqnarray}
\left|\langle\alpha\vert\beta\rangle\right|^{2} & = & e^{-\left|\alpha-\beta\right|^{2}}.
\end{eqnarray}
Defining the non-orthogonal basis
\begin{eqnarray}
\vert\varPsi_{1}\rangle & \equiv & \vert\alpha\sqrt{\eta},0\rangle_{12},\;\vert\varPsi_{2}\rangle\equiv\vert0,\alpha\sqrt{\eta}\rangle_{12},
\end{eqnarray}
the density matrix $\sigma_{{\rm ECS}}$ can be written as
\begin{eqnarray}
\sigma_{{\rm ECS}} & = & \mathcal{N}^{2}\!\left(\begin{array}{cc}
1 & e^{-\left(1-\eta\right)\left|\alpha\right|^{2}}\\
e^{-\left(1-\eta\right)\left|\alpha\right|^{2}} & 1
\end{array}\right).\label{eq:sigma_ECS}
\end{eqnarray}

\subsection{Derivation of Eq.~\eqref{eq:lossy_ECS_diagonal}}

\makeatletter
\renewcommand{\theequation}{B\arabic{equation}}
\makeatother
\setcounter{equation}{0}
Since the basis $\left\{ \vert\varPsi_{1}\rangle,\vert\varPsi_{2}\rangle\right\} $
form a non-orthogonal basis with overlap
\begin{eqnarray}
p & \equiv & \langle\varPsi_{1}\vert\varPsi_{2}\rangle=e^{-\eta\left|\alpha\right|^{2}},
\end{eqnarray}
we use the Gram-Schmidt procedure \citep{nielsen_chuang_2000book}
to construct an orthogonal basis

\begin{equation}
\left|\varPhi_{1}\right\rangle =\left|\varPsi_{1}\right\rangle ,\;\left|\varPhi_{2}\right\rangle =\frac{1}{\sqrt{1-p^{2}}}\!\left(\left|\varPsi_{2}\right\rangle -p\left|\varPsi_{1}\right\rangle \right).
\end{equation}
In this basis, the density matrix $\sigma_{{\rm ECS}}$ becomes
\begin{equation}
\sigma_{{\rm ECS}}=\mathcal{N}^{2}\!\left(\begin{array}{cc}
\!1+2pp_{\perp}+p^{2} & \!\left(p+p_{\perp}\right)\!\sqrt{1\!-\!p^{2}}\\
\!\left(p+p_{\perp}\right)\!\sqrt{1\!-\!p^{2}} & \!\!\left(1-p^{2}\right)p_{\perp}
\end{array}\right)\!,\label{eq:Matrix_basisO}
\end{equation}
with $p_{\bot}\!\equiv\!e^{-(1-\eta)\left|\alpha\right|^{2}}$. Diagonalizing
$\sigma_{{\rm ECS}}$, the spectral decomposition reads
\begin{eqnarray}
\sigma_{{\rm ECS}} & = & \gamma_{+}\vert\gamma_{+}\rangle\langle\gamma_{+}\vert+\gamma_{-}\vert\gamma_{-}\rangle\langle\gamma_{-}\vert,\label{eq:lossy_ECS_diagonal-1}
\end{eqnarray}
with eigenvalues
\begin{eqnarray}
\gamma_{\pm} & = & \frac{1}{2}\left(1\pm\sqrt{1-\det\sigma_{{\rm ECS}}}\right).
\end{eqnarray}
The corresponding eigenstates are
\begin{eqnarray}
\left|\gamma_{\pm}\right\rangle  & = & \left(\begin{array}{cc}
\pm\zeta_{\pm}e^{i\vartheta}, & \zeta_{\mp}\end{array}\right)^{\mathrm{T}},\label{eq:orthogonal-basis}
\end{eqnarray}
where $\zeta_{\pm}$ are defined in the main text. When transformed
into the original non-orthogonal basis $\left\{ \vert\varPsi_{1}\rangle,\vert\varPsi_{2}\rangle\right\} $,
these eigenstates can be expressed as
\begin{equation}
\left|\gamma_{\pm}\right\rangle =\mathcal{C}_{\pm}\left|\varPsi_{1}\right\rangle +\mathcal{D}_{\mp}\left|\varPsi_{2}\right\rangle ,\label{eq:bassi_nonorthogonal}
\end{equation}
with
\begin{eqnarray}
\mathcal{C}_{\pm} & = & \pm\zeta_{\pm}-p\mathcal{D}_{\mp},\;\mathcal{D}_{\mp}=\frac{\zeta_{\mp}}{\sqrt{1-p^{2}}}.
\end{eqnarray}

\subsection{Derivation of Eq.~\eqref{eq:QFI_nonorthogonal}}

\makeatletter
\renewcommand{\theequation}{C\arabic{equation}}
\makeatother
\setcounter{equation}{0}
Consider a general $d\times d$ density matrix $\rho$
with spectral decomposition
\begin{eqnarray}
\rho & = & \sum_{i=1}^{d}p_{i}\left|\psi_{i}\right\rangle \left\langle \psi_{i}\right|.
\end{eqnarray}
Assuming the unknown parameter $\phi$ to be encoded via the unitary
operator $U_{\phi}=e^{-iG\phi}$, where $G$ is the generator, the
resulting parameter-dependent state is
\begin{eqnarray}
\rho_{\phi} & = & U_{\phi}\rho U_{\phi}^{\dagger}=\sum_{i}p_{i}\left|\psi_{i}\left(\phi\right)\right\rangle \left\langle \psi_{i}\left(\phi\right)\right|,
\end{eqnarray}
where $\left|\psi_{i}\left(\phi\right)\right\rangle \!=\!U_{\phi}\left|\psi_{i}\right\rangle $.
The QFI can be computed by \citep{Zhong2013PRA,Jing2014bCTP}
\begin{eqnarray}
F & = & 4\sum_{i=1}^{d}p_{i}\Delta^{2}G_{ii}-\sum_{i\neq j}\frac{8p_{i}p_{j}}{p_{i}+p_{j}}\left|G_{ij}\right|^{2},\qquad
\end{eqnarray}
where
\begin{eqnarray}
\Delta^{2}G_{ii} & = & \left\langle \psi_{i}\right|G^{2}\left|\psi_{i}\right\rangle -\left\langle \psi_{i}\right|G\left|\psi_{i}\right\rangle ^{2},\\
G_{ij} & = & \left\langle \psi_{i}\right|G\left|\psi_{j}\right\rangle .
\end{eqnarray}
For the two-dimensional case of Eq.~\eqref{eq:lossy_ECS_diagonal-1},
the QFI simplifies to
\begin{eqnarray}
F & = & 4\left(\lambda_{+}\Delta^{2}G_{+}+\lambda_{-}\Delta^{2}G_{-}-4\lambda_{+}\lambda_{-}\!\left|G_{+-}\right|^{2}\right),\quad\label{eq:QFI_nonorthogonal-1}
\end{eqnarray}
where
\begin{eqnarray}
\Delta^{2}G_{\pm} & = & \left\langle \lambda_{\pm}\right|G^{2}\left|\lambda_{\pm}\right\rangle -\left\langle \lambda_{\pm}\right|G\left|\lambda_{\pm}\right\rangle ^{2},\\
G_{+-} & = & \left\langle \lambda_{+}\right|G\left|\lambda_{-}\right\rangle .
\end{eqnarray}

\bibliographystyle{apsrev4-1}
\bibliography{../../../ZW}

\end{document}